\documentclass[twocolumn,preprintnumbers,showpacs,amsmath,amssymb,superscriptaddress]{revtex4-1}

\usepackage{graphicx}
\usepackage{dcolumn}
\usepackage{color}
\usepackage{bm}
\usepackage{amsmath}
\usepackage{epstopdf}

\usepackage{dcolumn}

\begin{document}

\title{Absolute self-calibration of single-photon and multiplexed photon-number-resolving detectors}

\author{Lior Cohen}
\affiliation{%
Racah Institute of Physics, Hebrew University of Jerusalem, Jerusalem 91904, Israel}
\author{Yehuda Pilnyak}
\affiliation{%
Racah Institute of Physics, Hebrew University of Jerusalem, Jerusalem 91904, Israel}
\author{Daniel Istrati}
\affiliation{%
Racah Institute of Physics, Hebrew University of Jerusalem, Jerusalem 91904, Israel}
\author{Nicholas M. Studer}
\affiliation{%
Hearne Institute for Theoretical Physics, and Department of Physics and Astronomy, Louisiana State University, Baton Rouge, Louisiana 70776, United States}
\author{Jonathan P. Dowling}
\affiliation{%
Hearne Institute for Theoretical Physics, and Department of Physics and Astronomy, Louisiana State University, Baton Rouge, Louisiana 70776, United States}
\affiliation{%
NYU-ECNU Institute of Physics at NYU Shanghai, 3663 Zhongshan Road North, Shanghai, 200062, China}
\author{Hagai S. Eisenberg}
\affiliation{%
Racah Institute of Physics, Hebrew University of Jerusalem, Jerusalem 91904, Israel}


\begin{abstract}
Single-photon detectors are widely used in modern quantum optics experiments and applications. Like all detectors, it is important for these devices to be accurately calibrated. A single-photon detector is calibrated by determining its detection efficiency; the standard method to measure this quantity requires comparison to another detector. Here, we suggest a method to measure the detection efficiency of a single photon detector without requiring an external reference detector. Our method is valid for individual single-photon detectors as well as multiplexed detectors, which are known to be photon number resolving. The method exploits the photon-number correlations of a nonlinear source, as well as the nonlinear loss of a single photon detector that occurs when multiple photons are detected simultaneously. We have analytically modeled multiplexed detectors and used the results to experimentally demonstrate calibration of a single photon detector without the need for an external reference detector.
\end{abstract}


\maketitle

Information about the photon number is required for applications in many diverse fields such as linear-optical quantum computing\cite{Knill,Kok}, super-resolution\cite{Cohen}, supersensitive microscopy\cite{Israel}, foundations of quantum mechanics\cite{Gerry}, and quantum key distribution\cite{Horikiri}. To date, there are a few techniques to measure photon-number\cite{Hadfield}. One leading approach is to use single photon detectors (SPDs) with splitters to separate photons into different SPDs (spatial multiplexing) or as an array of SPDs\cite{Dolgoshein}. It is also known that one SPD can be used if the splitters separate the photons into different time slots (time multiplexing)\cite{Fitch03}. The signal from all SPDs is summed yielding the photon-number information. Though providing this information, these multiplexing techniques are not considered as full photon-number-resolving (PNR) detectors due to saturation of the elements\cite{Migdall}.
We use 'multiplexing PNR detectors' as a general name for time multiplexing, spatial multiplexing and array of SPDs throughout this paper.

To have precise photon number information of the measured quantum state, the detector must be characterized, and in particular its detection efficiency must be measured. Characterization of the detection efficiency is here considered to be calibration of the device.
The standard procedure to calibrate SPDs is to use correlated photon pairs from a twin-beam state. This method was first suggested in 1977 by Klyshko\cite{Klyshko} and demonstrated experimentally two years later\cite{Kitaeva}. Two detectors are required for this method because the coincidence rate must be known to calculate the efficiencies.
Recently, the method was adapted to PNR detectors\cite{Perina, Worsley, Avella}.

There are other methods to calibrate the detection efficiency\cite{Chen, Lopez}. One such method utilizes a single SPD, which is time-multiplexed to temporally separate the incident photons\cite{Chen}. Then, the detection efficiency is found as in the two detector method. Here, we develop a model to characterize a multiplexing PNR detector, and apply the analysis to calibration of a single SPD. Using the photon statistics of the single-mode and two-mode squeezed vacuum states (SMSV and TMSV, respectively), we show how the detection efficiency can be found without a reference detector or multiplexing.


This paper is arranged as follows: a model for a general multiplexing PNR detector is presented in Sec. \ref{Model}. In Sec. \ref{Setup} we limit the discussion to one SPD and show how the efficiency can be measured using SMSV and TMSV light. The setup to perform this calibration is also described there. Results of the calibration procedures are presented in Sec. \ref{Results}. There we also compare between the use of SMSV and TMSV for the calibration.

\section{The characterization model}\label{Model}

Given a multiplexing PNR detector, there is a problem with counting the photon number due to several internal effects distorting the measurement statistics\cite{Dovrat}. The incident photon statistics can however be reconstructed if the distortion effects are well quantified. We consider several detection parameters: efficiency, number of SPD elements (finite detector size), dark count rate, and cross-talk rate. To date, there lacks an analytical model for all of these effects. In particular, the combined effects of finite-size with cross-talk are not well known\cite{DovratSim}. Now, we present an analytical model which incorporates all of these effects.

\subsection{Loss}\label{sec:loss}

When a photon hits the detector, there is a non-zero probability that either the avalanche process will not start or will stop before a detection occurs\cite{saleh}. This is an intrinsic property of any realistic device, but can also be attributed to inefficient light coupling to the device. In such a scenario the photon is considered lost. The detection efficiency is then defined as the probability for detecting a single photon. We assume the detection efficiency is uniform for all SPD elements and is denoted by $\eta$. If $n$ photons hit the detector, the probability for $m$ elements to be activated is given by a binomial distribution\cite{Dovrat}:

\begin{eqnarray}\label{loss}
M^{\eta}_{loss}(m,n) =\binom{n}{m}\eta^m(1-\eta)^{n-m}\,,
\end{eqnarray}
where $\binom{n}{m}=\frac{n!}{m!(n-m)!}$ for $n\geq m \geq 0$ or zero otherwise.
Here we assume each photon hits a different element. This assumption is not valid in general but will be later corrected for by including the effects of finite detector size.

\subsection{Finite-size}

Each individual element of the multiplexing PNR detector is an SPD. As such, the signal from each element does not depend on the number of photons hitting it. Therefore, if more than one photon hits an element, only one can be detected, causing a non-linear loss of photons and a distortion of the incident photon statistics. The probability for $m$ photons to hit $k$ different elements at an $N$-element detector, is\cite{Paul96}:

\begin{eqnarray}\label{FS}
M^N_{FS}(k,m) =\frac{1}{N^m}\binom{N}{k}k!S(m,k)\,,
\end{eqnarray}
where $ S(m,k) = \frac{1}{k!}\sum_{j=0}^k(-1)^{k-j}\binom{k}{j} j^m $ are known as the Stirling numbers of the second kind\cite{wolfram}.
 	
\subsection{Dark-counts}

After $k$ elements fire due to photon detections, there are still $(N-k)$ elements which are free to be activated due to a dark-count \--- a false event without a photon hit. This is typically due to thermal electrons. We assume that each element has equal probability $d$ for this event to occur. We define $p$ to be the total number of elements that fire, including those elements which report a dark count. Then $p-k$ is the number of elements that fire due to a dark count event. The probability for $(p-k)$ elements to be activated due to dark-counts where $(N-k)$ elements are available is

\begin{eqnarray}\label{DC}
M^{d}_{DC}(p,k) =\binom{N-k}{p-k}d^{(p-k)}(1-d)^{N-p}\,.
\end{eqnarray}
 	
\subsection{Cross-talk}

Cross-talk is an effect where a recombination of an electron and a hole generates a photon and this photon is detected in a non-activated neighbor element\cite{Buzhan}.
All multiplexing PNR detectors suffer from this effect but, it is not relevant when the SPDs are distant and the cross-talk counts can be temporally filtered. Where the detector is a SPD array, the cross-talk counts cannot be filtered and the cross-talk effect is relevant.
Cross-talk is most likely to happen at nearest neighbor element and we neglect other scenarios.

Up to date, a few cross-talk models are available\cite{Eraerds, Akiba, Afek, Dovrat}. Each has its own advantages and disadvantages, but none of them takes into account the finite size of the detector. Thus, we introduce a new model for the cross-talk:   
until this point, uniformity was the only assumption. Yet, in order to solve analytically the cross-talk effect, more assumptions must be made. The probability of cross-talk strongly dependents on the number of non-activated neighbors, but it is impossible to know how many nearest neighbors are available. Instead, we check how many nearest neighbors are available on average, where $p$ elements already were activated, and we plug this number in as the total effective number of nearest neighbors, $\rm{ENN}=4(1-\frac{p}{N})\frac{N-\sqrt{N}}{N-1}$.  This linearly dependent formula is reasonable as it is zero if all elements are not available $(p=N)$. On the other limit, if all elements are available the $\rm{ENN}$ nears four, a limit imposed by the rectangular detector's edge. We derive this formula simply by randomly choosing $p$ elements and counting their nearest neighbors, and then averaging many different configurations.

We define $x$ to be the probability for cross-talk to one of the available nearest neighbors. We neglect terms proportional to higher powers of $x$ by assuming $x \ll 1$. In particular, we neglect cross-talk generated by another cross-talk and more than one cross-talk event  per element. Thus, the probability for one element to generate a cross-talk event is $4x(1-\frac{p}{N})\frac{N-\sqrt{N}}{N-1}$ and for $p$ elements to generate $\ell$ cross-talk events is just a binomial combination. Thus, under the mentioned assumptions, the probability for $(s-p)$ elements to be activated by cross-talks from $p$ elements is,

\begin{eqnarray}\label{XT}
\nonumber& M^{\tilde x}_{XT}(s,p) =\binom{p}{s-p}\left(\tilde x\left(1-\frac{p}{N}\right)\right)^{s-p}\left(1-\tilde x\left(1-\frac{p}{N}\right)\right)^{2p-s}\\
\end{eqnarray}
where we define $\tilde x = 4x\frac{N-\sqrt{N}}{N-1}$.

\subsection{The detected probabilities}

The real photon number probabilities ($\vec P_{\rm{real}}$) is related to the detected photon number probabilities ($\vec P_{\rm{det}}$) by

\begin{eqnarray}\label{real}
 \vec P_{\rm{det}} = \mathbf{M}_{XT}\cdot\mathbf{M}_{D}\cdot \mathbf{M}_{FS}\cdot \mathbf{M}_{L}\cdot \vec P_{\rm{real}}\,,
\end{eqnarray}
where $\mathbf{M}_{XT},\mathbf{M}_{D},\mathbf{M}_{FS},\mathbf{M}_{L}$ are matrices quantifying the cross-talk, dark-counts, finite-size and loss  effects, respectively. The ordering of the loss and finite size matrices is important here, but we do not show a proof of the correct ordering here. Instead, we observe that Eq. \ref{real} agrees with previous theoretical results that do not take matrix ordering into account\cite{Fitch03,Paul96}.

We first calculate the detected statistics for an $n$-photon Fock state, a state with a fixed number of $n$ photons; we can then generalize to any other state by averaging the results over the real photon statistics. The probability for $s$ detection events to occur due to an incident $n$-photon Fock state after accounting for all the distorting effects is,

\begin{eqnarray}\label{ProbTOT}
&& P^{n}_{\rm{det}}(s|\eta,d,N,\tilde{x}) = \\
\nonumber &&\sum_{p=0}^s\binom{p}{s-p}\left(\tilde x\left(1-\frac{p}{N}\right)\right)^{(s-p)}\left(1-\tilde x\left(1-\frac{p}{N}\right)\right)^{2p-s} \\
\nonumber && \binom{N}{p}\sum_{j=0}^p\binom{p}{j}(-1)^{p-j}(1-d)^{N-j}\left(1-\eta+\frac{j\eta}{N}\right)^n\,.
\end{eqnarray}
This result is proven in appendix \ref{appTH1} and agrees with previous analytical results substituting $\tilde{x}=0, d=0$\cite{Paul96} and $\eta=1$\cite{Fitch03}.
The agreement between the results shows that the loss should be operated before the finite-size effect as mentioned above.

\section{Experimental setup for SPD calibration}\label{Setup}

From this point we focus on a single SPD, i.e. for $N=1$. In this case, there are only two possibilities; there is either a detection event or not. Mathematically it means $s=0$ or $1$ in Eq. \ref{ProbTOT}, and thus the cross-talk summation vanishes. 

\begin{figure}[t]
\centering
\fbox{\includegraphics[width=\linewidth]{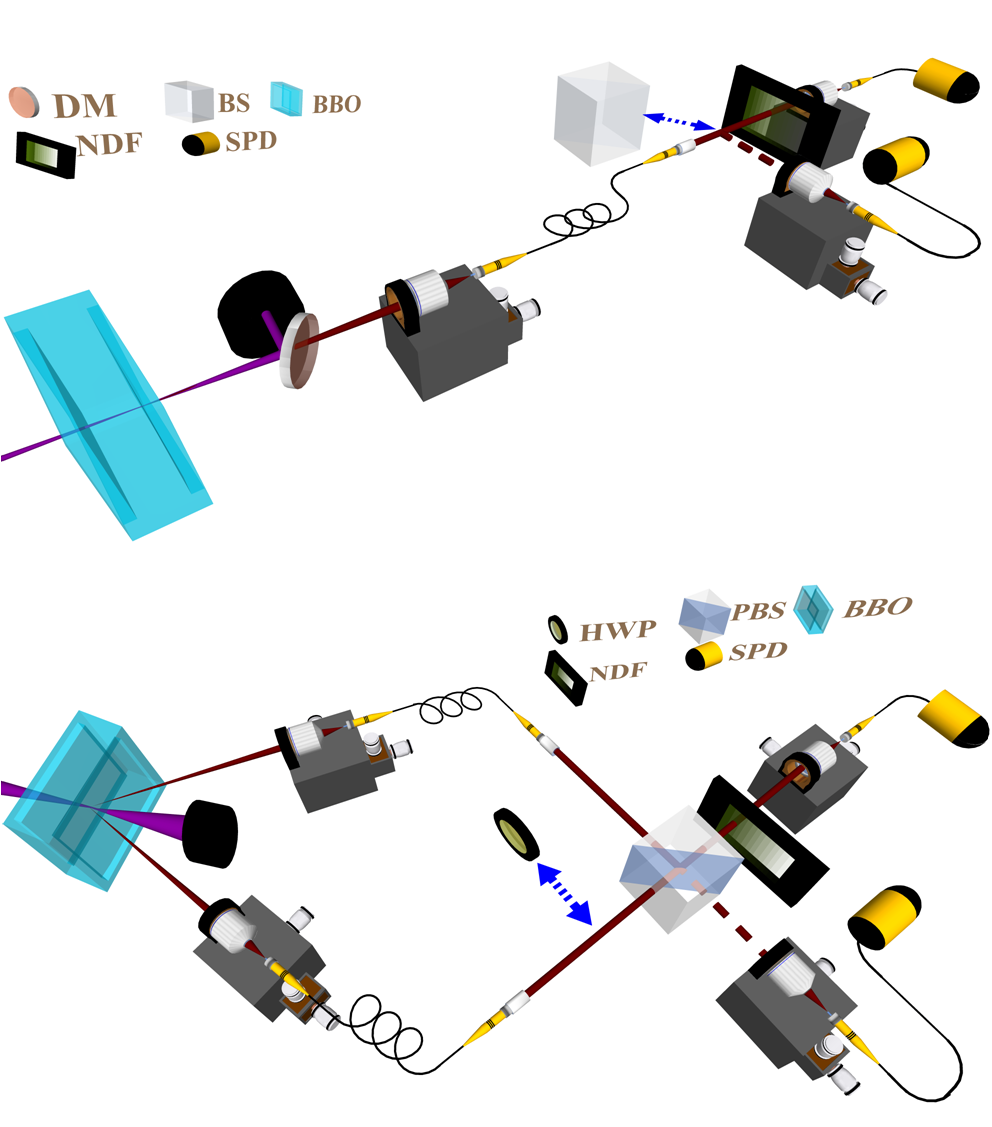}}
\caption{The experimental setup. The upper part is the setup where a SMSV state is used and the lower part where a TMSV state is used. DM - dichroic mirror, PBS - polarizing beam splitter, HWP - half wave-plate, NDF - variable neutral density filter, BD - beam dump. Note that SPD2 is only for comparison to the two detector calibration procedure.}
\label{fig1}
\end{figure}

In order to calibrate the detector, we have used both single-mode squeezed vacuum (SMSV) and two-mode squeezed vacuum (TMSV) states with a calibrated adjustable attenuator on the SV light. Although only one of these states must be used, we show that both schemes will work for calibration purposes. The attenuated SV light is directed towards a bandpass filter and then sent into a single mode fiber coupled into the detector. If a TMSV state is used, the two modes are fixed to orthogonal polarizations and spatially combined with a polarizing beam splitter (PBS) (see Fig. \ref{fig1}).

We found it convenient to define the odds $O_{\rm{det}}^{\rm{n}}(\eta,d,1,0) \equiv \frac{P_{\rm{det}}^{\rm{n}}(s=1|\eta,d,1,0)} {P_{\rm{det}}^{\rm{n}}(s=0|\eta,d,1,0)}$ of a detection event. We also replace $\eta \rightarrow \eta t $, where $t$ is the transmission of the calibrated adjustable attenuator, and henceforth $\eta$ is the fixed efficiency and $t$ is a variable.
Following these changes, Eq. \ref{ProbTOT} is now reduced to:
\begin{align}
O_{\rm{det}}^{\rm{SMSV}}(\eta t,d,1,0) = \left(\frac{\sqrt{1+(2-\eta t)\eta \bar{n}t}}{1-d}-1\right) \approx \frac{(1-\frac{\eta t}{2})\eta \bar{n}t+d}{1-d}\,, \label{ProbRatioSMSV}
\end{align}
\begin{align}
&O_{\rm{det}}^{\rm{TMSV}}(\eta t,d,1,0) =  \frac{(1-\frac{\eta t}{2})\eta \bar{n}t+d}{1-d} \label{ProbRatioTMSV}
\,.
\end{align}
Here Eqs. \ref{ProbRatioSMSV} and \ref{ProbRatioTMSV} are for SMSV and TMSV states, respectively.
The approximation is the Taylor expansion for $\bar{n}\ll1$ and the full derivation is found in appendix \ref{appTH3}.

Experimentally, the probability of detection is given by the ratio of the number of detection events to the number of pump pulses. This probability is measured while varying the transmission of the Neutral Density Filter (NDF). The efficiency parameter is then extracted from a second-order fit to Eq. \ref{ProbRatioSMSV} or Eq. \ref{ProbRatioTMSV}. We can also calibrate the multiplexed PNRs in a similar manner, though we do not demonstrate it in this manuscript.

The experimental setup is described in Fig. \ref{fig1}.
780\,nm photon pairs are generated by a spontaneous parametric down-conversion (SPDC) process from a 2\,mm thick $\beta$-BaB$_2$O$_4$ (BBO) crystal using a 390\,nm doubled Ti:Sapphire pulsed laser.
In the first part, we have used a collinear type-I SPDC to generate a horizontally polarized SMSV state. In the second part, we have used a non-collinear type-II SPDC to generate a TMSV state. The two modes have been set in orthogonal polarization modes, and spatially overlapped by a PBS. For comparison, another SPD has been used to measure the detection efficiency in the two detector method. For clarity, the beam path to the reference detector is indicated by a dashed line, where a beamsplitter (BS) or half-wave plate (HWP) were added to divert photons to the reference detector. We note that we have used a calibrated NDF as a convenient method for a known attenuation. Any other self-calibrated attenuation method can be used; for instance, the attenuation of SMSV can be applied by a single rotating polarizer.   

\section{Experimental results}\label{Results}

The present scheme is useful for evaluating the detection efficiency of a SPD, due to the unique photon statistics of the SV light and the non-linear loss of the SPD. The non-linear loss alters the linear dependency of the single counts on the attenuation and the detection efficiency is extracted from the curvature.

\begin{figure}[t]
\centering
\fbox{\includegraphics[width=\linewidth]{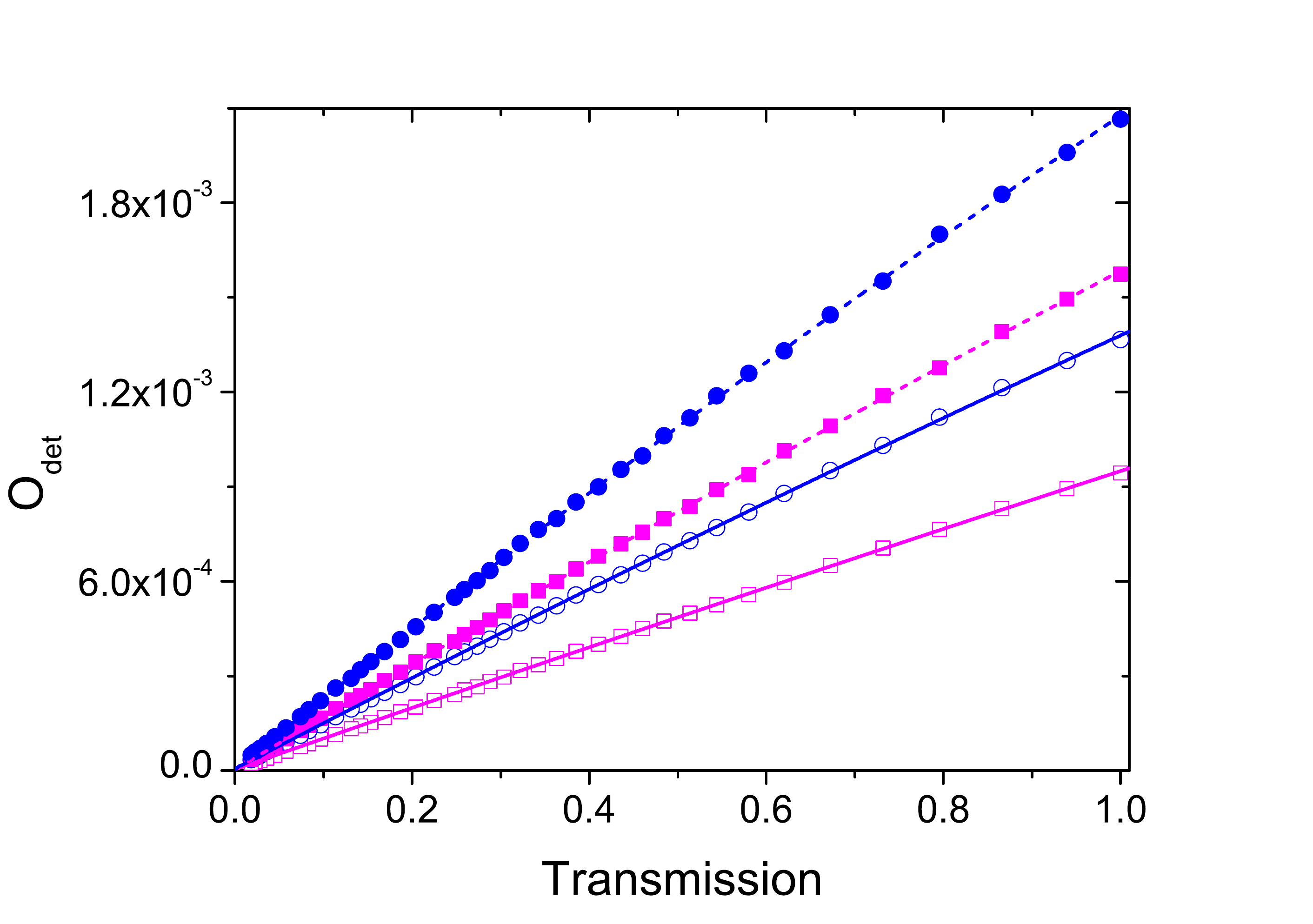}}
\caption{(Color Online) The odds of a detection event as a function of the NDF transmission for two separate detectors.
Solid  and empty symbols denote data from using TMSV and SMSV, respectively. Solid and dashed lines are fits to Eqs. \ref{ProbRatioSMSV} and \ref{ProbRatioTMSV}, respectively. SPD\#1 is represented in blue circles and SPD\#2 is represented in pink boxes. Error bars are assumed to be due to Poissonian noise and are smaller than the symbol sizes, thus they are not displayed.}
\label{fig2}
\end{figure}

The SPD counts were accumulated for one second for a range of $40$ different attenuation values of the NDF. The probability for a photon detection is measured by the single counts divided by the total number of experiment runs. We repeat the experiment for two separate SPDs using both SMSV and TMSV in order to demonstrate the ability to calibrate detectors of different efficiency. The results are presented in Fig. \ref{fig2}. In each of the four measurements the data is fit to a second-order polynomial, i.e. $a_2t^2+a_1t+a_0$. According to Eqs. \ref{ProbRatioSMSV} and \ref{ProbRatioTMSV} the efficiency is $\eta = -2\frac{a_2}{a_1}$.

In table \ref{table} the results for the efficiency calibration by the presented single detector method are summarized. Those results are compared to the two detector method showing good agreement between the two methods, for all used detectors and for both experimental setups. The detection efficiency is lower in the SMSV setup due to weaker coupling to single-mode fiber, which is probably caused by spatial walk-off inside the non-linear crystal. This inefficient coupling is a loss factor well observed by both calibration methods.

\begin{table}[b]
\centering
\caption{The efficiencies measured by the presented single detector method ($\eta_1$) and the two detector method ($\eta_2$). Note that the SMSV efficiencies are lower than in the TMSV case due to weaker coupling into the single-mode fiber.}
\begin{tabular}{cccc}
\hline
SPD $\#$ & SV light & $\eta_1$ & $\eta_2$ \\
\hline
$1$ & SMSV & $11.3\pm1.1\%$ & $11.8\pm0.9\%$ \\
$2$ & SMSV & $7.4\pm0.9\%$ & $8.1\pm0.9\%$ \\
$1$ & TMSV & $17.4\pm1.0\%$ & $17.3\pm0.8\%$ \\
$2$ & TMSV & $12.7\pm0.9\%$ & $11.7\pm0.8\%$ \\
\hline
\end{tabular}
  \label{table}
\end{table}

\begin{figure}[t]
\centering
\fbox{\includegraphics[width=\linewidth]{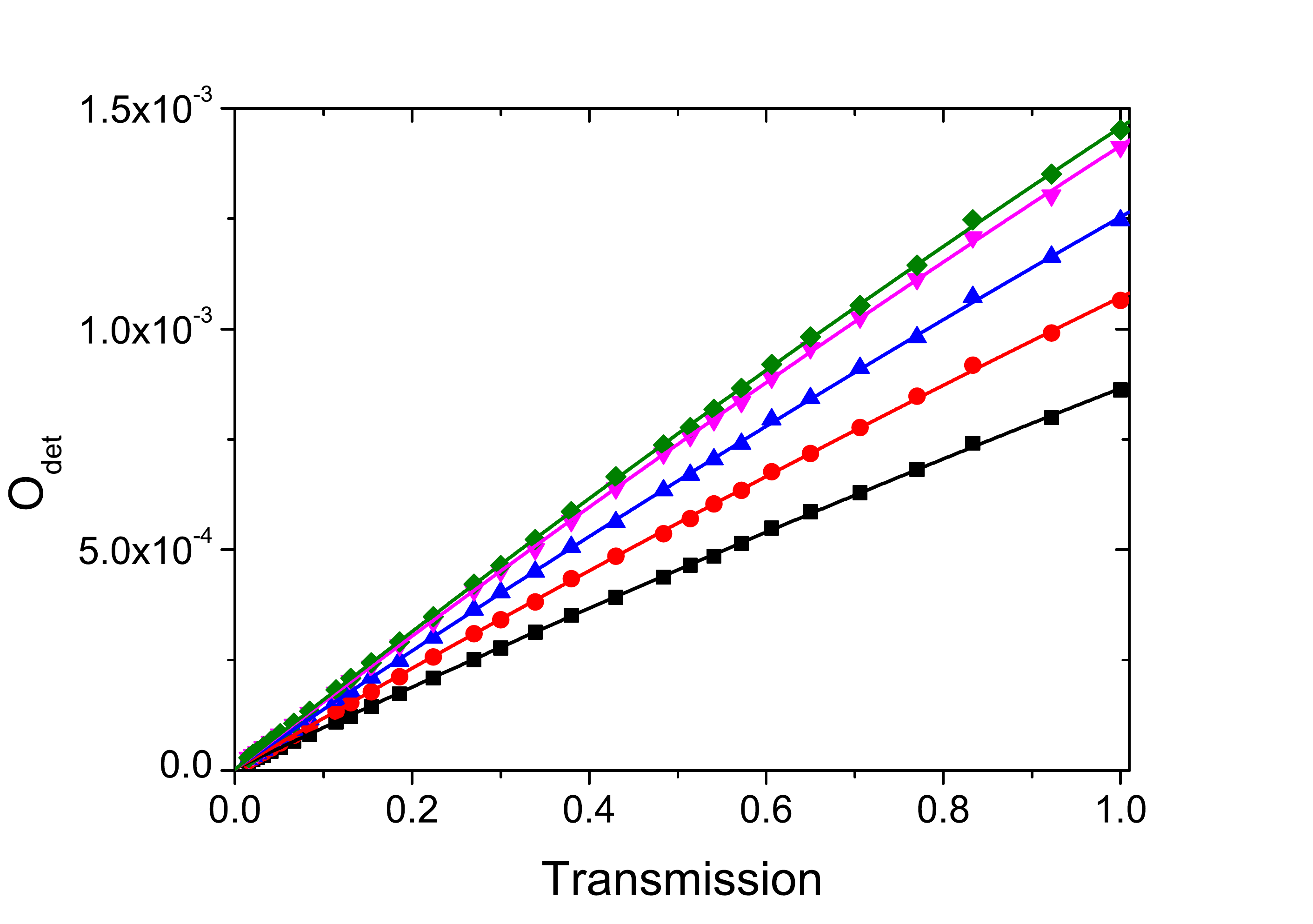}}
\caption{(Color Online) The odds of a detection event as a function of the NDF transmission for SPD \#1 when the pump power is varied.
Green diamonds are for pump power of 250 mW, pink downward triangles for 240 mW, blue triangles for 215 mW, red circles for 180 mW and black boxes for 145 mW. Error bars are assumed to be due to Poissonian noise and are smaller than the symbol sizes, thus they are not displayed.}
\label{fig3}
\end{figure}

In order to show the presented method is valid for any pump intensity, we repeated the experiment using TMSV and SPD \#1 for different pump intensities. The results of this process are shown in Fig. \ref{fig3}. As before, we fit the measurements to a second-order polynomial and the efficiency is calculated from the polynomial coefficients. 

The results for different pump powers are summarized in table \ref{table2}. A good agreement is shown between different pump powers, where a standard deviation of $0.5\%$ is found. The standard deviation consists with the error values of the detection efficiency which were calculated separately. 

\begin{table}[htbp]
\centering
\caption{The efficiencies as measured by the single detector method ($\eta_1$) with SPD1 and TMSV light for different pump powers.}
\begin{tabular}{cc}
\hline
Pump power & $\eta_1$ \\
\hline
$145\,\rm{mW}$ & $17.9\pm0.8\%$ \\
$180\,\rm{mW}$ & $16.5\pm0.9\%$ \\
$215\,\rm{mW}$ & $17.2\pm0.8\%$ \\
$240\,\rm{mW}$ & $16.8\pm0.7\%$ \\
$250\,\rm{mW}$ & $17.6\pm0.7\%$ \\
\hline
\end{tabular}
  \label{table2}
\end{table}

\section{Summary}\label{Summary}

We have presented a model to characterize  a PNR detector based on SPDs. This model predicts the detected photon statistics in the presence of loss, finite size, dark counts and cross-talk. The model is valid also for a single SPD. The predicted statistics show an efficiency dependence when detecting SV light. Thus, the efficiency can be measured without a reference detector. We have experimentally measured the efficiency and successfully compared it to the two detector method.

\section{Acknowledgements}
JPD and NMS would like to acknowledge support from the Army Research Office, the Defense Advanced Research Project Agency, and the Louisiana Economic Development Assistantship Program.

\section{appendix}

\subsection{Probability calculation}\label{appTH1}
We first replace the matrix products in Eq. \ref{real} into summations and substitute the matrix values according to Eqs. \ref{loss} - \ref{XT}:

\begin{eqnarray}\label{ProbTOT2}
&& P^{n}_{\rm{det}}(s|\eta,d,N,\tilde{x}) =\\
\nonumber &&\sum_{p=0}^N\binom{p}{s-p}\left(\tilde x\left(1-\frac{p}{N}\right)\right)^{s-p}\left(1-\tilde x\left(1-\frac{p}{N}\right)\right)^{2p-s}\\
\nonumber &&\sum_{k=0}^N\binom{N-k}{p-k}d^{(p-k)}(1-d)^{N-p}\\
\nonumber && \sum_{m=0}^N\frac{1}{N^m}\binom{N}{k}k!S(m,k)\binom{n}{m}\eta^m(1-\eta)^{n-m}\,.
\end{eqnarray}
Now, we focus on the two last lines in Eq. \ref{ProbTOT2}. We notice that $m\leq{n}$ and $k\leq{p}\,$, because the loss effect cannot increase the photon number and dark-counts cannot decrease it. After reorder the summations and substitute $\binom{N}{k}\binom{N-k}{p-k} = \binom{N}{p}\binom{p}{k}$ we get:
\begin{eqnarray}\label{ProbTOT3}
&& \binom{N}{p} \sum_{m=0}^n \frac{1}{N^m} \binom{n}{m}\eta^m(1-\eta)^{n-m}\\
\nonumber&& \sum_{k=0}^p\binom{p}{k}d^{(p-k)}(1-d)^{N-p}\sum_{j=0}^k(-1)^{k-j}\binom{k}{j}j^m\,.
\end{eqnarray}
We reorder the summations in the second line, use $\binom{p}{k}\binom{k}{j} = \binom{p}{j}\binom{p-j}{k-j} $ and replace the summation index $k\rightarrow k-j$, resulting in the second line to be:
\begin{eqnarray}\label{ProbTOT4}
\sum_{j=0}^p\binom{p}{j}j^m\sum_{k=0}^{p-j}\binom{p-j}{k}(-1)^k d^{p-k-j}(1-d)^{N-p}\,.
\end{eqnarray}
The inner summation equals to $(1-d)^{N-j}(-1)^{p-j} $. Substituting this in Eq. \ref{ProbTOT3} and reordering the summations, we get:
\begin{eqnarray}\label{ProbTOT5}
\nonumber&\binom{N}{p}\sum_{j=0}^p\binom{p}{j}(1-d)^{N-j}(-1)^{p-j}
\sum_{m=0}^n\binom{n}{m} \left(\frac{\eta j}{N}\right)^m (1-\eta)^{n-m}\,.\\
\end{eqnarray}
However the second summation is a binomial expansion and regrouping it restores the third line of Eq. \ref{ProbTOT}.
The second line remains almost as is. The only change is that we change the upper limit to $s$, the number of activated elements by signal photon and dark counts, as we neglect higher-order cross-talks, which means the number of cross-talks is limited by the number of already activated elements. The accurate upper limit is $\min(p,N-p)$, but we checked it numerically and the difference is negligible.

\subsection{Calculating SPD probabilities for SV states}\label{appTH3}

Substituting $N=1$ in Eq. \ref{ProbTOT}, i.e.:

\begin{align}\label{ProbRatio2}
 P^{n}_{\rm{det}}(s|\eta,d,1,\tilde{x}) =\sum_{j=0}^p\binom{p}{j}(-1)^{p-j}(1-d)^{1-j}\left(1-\eta+j\eta\right)^n\,.
\end{align}
The first summation in Eq. \ref{ProbTOT} vanishes as a SPD has no neighbors to cross-talk to. We write the probabilities for no detection and for one photon detection explicitly:

\begin{eqnarray}
&& P^{n}_{\rm{det}}(0|\eta,d,1,0)=(1-d)\left(1-\eta\right)^n\,, \\
\label{P0_Fock}
&&P^{n}_{\rm{det}}(1|\eta,d,1,0)=1-(1-d)\left(1-\eta\right)^n\,.
\label{P1_Fock}
\end{eqnarray}
The probability to have zero photon counts is just the probability to not detect $n$ photons times the probability to not have dark-counts. The probability to get one photon detection is just the complementary probability.
Next, we average over the photon statistics of the SV state.

For a TMSV state any mode has  photon statistics of $P(n)=(1-x)x^n$ where $x$ is related to $\bar{n}$, the average photon number, by $x=\frac{\bar{n}}{1+\bar{n}} $. After combining the two modes spatially the probability for $2n$-photons is $P_{TMSV}(2n) = (1-x)x^n$. We average on the statistics and get:
\begin{eqnarray}
& P^{\rm{TMSV}}_{\rm{det}}(0|\eta,d,1,0)=\frac{(1-d)(1-x)}{1-x(1-\eta)^2}\,,\\
\label{P0_TMSV}
& P^{\rm{TMSV}}_{\rm{det}}(1|\eta,d,1,0)=1-\frac{(1-d)(1-x)}{1-x(1-\eta)^2}\,.
\label{P1_TMSV}
\end{eqnarray}
Taking the ratio of the two last equations gives Eq. \ref{ProbRatioTMSV}.

For a SMSV state the photon statistics is $P_{SMSV}(n)=\cos^2{\frac{n\pi}{2}}\frac{n!}{2^n((\frac{n}{2})!)^2}\frac{\tanh{r}^n}{\cosh{r}}$, where $r$ is the squeezed parameter. After the averaging we get:
\begin{eqnarray}
& P^{\rm{SMSV}}_{\rm{det}}(0|\eta,d,1,0)=(1-d)\frac{1}{\sqrt{1+(2\eta-\eta^2)\bar{n}}}\,,\\
\label{P0_SMSV}
& P^{\rm{SMSV}}_{\rm{det}}(1|\eta,d,1,0)=1-(1-d)\frac{1}{\sqrt{1+(2\eta-\eta^2)\bar{n}}}\,,
\label{P1_SMSV}
\end{eqnarray}
where $\bar{n} = \sinh^2{r}$ is the average photon number and the hyperbolic function identities, $ \cosh{\big(\tanh^{-1}{r}\big)}=\frac{1}{\sqrt{1-r^2}}\,,\,\cosh^2{r}-\sinh^2{r}=1 $, were used.
Taking the ratio of the two last equations yields Eq. \ref{ProbRatioSMSV}.


\begin{thebibliography}{0}%
\makeatletter
\providecommand \@ifxundefined [1]{%
 \@ifx{#1\undefined}
}%
\providecommand \@ifnum [1]{%
 \ifnum #1\expandafter \@firstoftwo
 \else \expandafter \@secondoftwo
 \fi
}%
\providecommand \@ifx [1]{%
 \ifx #1\expandafter \@firstoftwo
 \else \expandafter \@secondoftwo
 \fi
}%
\providecommand \natexlab [1]{#1}%
\providecommand \enquote  [1]{``#1''}%
\providecommand \bibnamefont  [1]{#1}%
\providecommand \bibfnamefont [1]{#1}%
\providecommand \citenamefont [1]{#1}%
\providecommand \href@noop [0]{\@secondoftwo}%
\providecommand \href [0]{\begingroup \@sanitize@url \@href}%
\providecommand \@href[1]{\@@startlink{#1}\@@href}%
\providecommand \@@href[1]{\endgroup#1\@@endlink}%
\providecommand \@sanitize@url [0]{\catcode `\\12\catcode `\$12\catcode
  `\&12\catcode `\#12\catcode `\^12\catcode `\_12\catcode `\%12\relax}%
\providecommand \@@startlink[1]{}%
\providecommand \@@endlink[0]{}%
\providecommand \url  [0]{\begingroup\@sanitize@url \@url }%
\providecommand \@url [1]{\endgroup\@href {#1}{\urlprefix }}%
\providecommand \urlprefix  [0]{URL }%
\providecommand \Eprint [0]{\href }%
\providecommand \doibase [0]{http://dx.doi.org/}%
\providecommand \selectlanguage [0]{\@gobble}%
\providecommand \bibinfo  [0]{\@secondoftwo}%
\providecommand \bibfield  [0]{\@secondoftwo}%
\providecommand \translation [1]{[#1]}%
\providecommand \BibitemOpen [0]{}%
\providecommand \bibitemStop [0]{}%
\providecommand \bibitemNoStop [0]{.\EOS\space}%
\providecommand \EOS [0]{\spacefactor3000\relax}%
\providecommand \BibitemShut  [1]{\csname bibitem#1\endcsname}%
\let\auto@bib@innerbib\@empty
\end{thebibliography}%


\begin{thebibliography}{1}

\bibitem{Knill} E. Knill, R. Laflamme, and G. J. Milburn, “A scheme for efficient quantum computation with linear optics,” \href{http://dx.doi.org/10.1103/RevModPhys.79.135}{\textit{Nature} (London) \textbf{409}, 46 (2001).}

\bibitem{Kok} P. Kok, W. J. Munro, K. Nemoto, T. C. Ralph, J. P. Dowling, and G. J. Milburn, “Linear optical quantum computing with photonic qubits,” \href{https://doi.org/10.1103/RevModPhys.79.135}{\textit{Rev. Mod. Phys.} \textbf{79}, 135 (2007).}

\bibitem{Cohen} L. Cohen, D. Istrati, L. Dovrat, and H. S. Eisenberg, “Super-resolved phase measurements at the shot noise limit by parity measurement,” \href{https://doi.org/10.1364/OE.22.011945} {\textit{Opt. Exp.} \textbf{22}, 11945 (2014).} 

\bibitem{Israel} Y. Israel, S. Rosen, and Y. Silberberg, “Supersensitive polarization microscopy using NOON states of light,”
 \href{https://doi.org/10.1103/PhysRevLett.112.103604} {\textit{Phys. Rev. Lett.} \textbf{112}, 103604 (2014).}

\bibitem{Gerry} C. C. Gerry, J. Mimih, and A. Benmoussa, “Maximally entangled coherent states and strong violations of Bell-type inequalities,” \href{https://doi.org/10.1103/PhysRevA.80.022111}{\textit{Phys. Rev. A} \textbf{80}, 022111 (2009).}

\bibitem{Horikiri} T. Horikiri and T. Kobayashi, “Decoy state quantum key distribution with a photon number resolved heralded single photon source,” \href{https://doi.org/10.1103/PhysRevA.73.032331}{\textit{Phys. Rev. A} \textbf{73}, 032331 (2006).}

\bibitem{Hadfield} R. H. Hadfield, “Single-photon detectors for optical quantum information applications,'' \href{http://www.nature.com/doifinder/10.1038/nphoton.2009.230}{ \textit{Nat. Photonics} \textbf{3}, 696 (2009).} 

\bibitem{Dolgoshein} B. Dolgoshein, V. Balagura, P. Buzhan, M. Danilov, L. Filatov, E. Garutti, M. Groll, A. Ilyin, V. Kantserov, V. Kaplin, A. Karakash, F.Kayumov, S. Klemin, V. Korbel, H. Meyer, R. Mizuk, V. Morgunov, E. Novikov, P. Pakhlov, E. Popova, V. Rusinov, F. Sefkow, E. Tarkovsky and I. Tikhomirov,
``Status report on silicon photomultiplier development and its applications,'' 
\href{https://doi.org/10.1016/j.nima.2006.02.193}{  \textit{Nucl. Instrum. Methods} 563 \textbf{2}, 368 (2006). } 

\bibitem{Fitch03} M. J. Fitch, B. C. Jacobs, T. B. Pittman, and J. D. Franson, ``Photon-number resolution using time-multiplexed single-photon detectors,'' \href{https://doi.org/10.1103/PhysRevA.68.043814}{\textit{Phys. Rev. A} \textbf{68}, 043814 (2003).}

\bibitem{Migdall} A. Migdall, S. V. Polyakov, J. Fan, and F. C. Bienfang, \textit{Single-Photon Generation and Detection: Physics and Applications} (Vol. 45). Academic Press. (2013).

\bibitem{Klyshko} D. N. Klyshko, ``Utilization of vacuum fluctuations as an optical brightness standard,'' \textit{Sov. J. Quantum Electron.} \textbf{7}, 591 (1977).

\bibitem{Kitaeva} G. Kitaeva, A. N. Penin, V. V. Fadeev, and Yu. A. Yanait, ``Measurement of brightness of light fluxes using vacuum fluctuations as a reference,'' \textit{Sov. Phys Dokl.} \textbf{24}, 564 (1979).

\bibitem{Perina} J. Pe\v{r}ina, Jr., O. Haderka, A. Allevi, and M. Bondani, `Absolute calibration of photon-number-resolving detectors with an analog output using twin beams,'' \href{http://dx.doi.org/10.1063/1.4863433}{\textit{Appl. Phys. Lett.} \textbf{104}, 041113 (2014).}

\bibitem{Worsley} A. P. Worsley, H. B. Coldenstrodt-Ronge, J. S. Lundeen, P. J. Mosley, B. J. Smith, G. Puentes, N. Thomas-Peter, and I. A.Walmsley, ``Absolute efficiency estimation of photon-number-resolving detectors using twin beams,'' \href{https://doi.org/10.1364/OE.17.004397}{\textit{Opt. Exp.} \textbf{17}, 4397 (2009).}

\bibitem{Avella} A. Avella, G. Brida, I. P. Degiovanni, M. Genovese, M. Gramegna, L. Lolli, E. Monticone, C. Portesi, M. Rajteri, M. L. Rastello, E. Taralli, P. Traina, and M. White, ``Self consistent, absolute calibration technique for photon number resolving detectors,'' \href{https://doi.org/10.1364/OE.19.023249}{\textit{Opt. Exp.} \textbf{19}, 23249 (2011).}

\bibitem{Chen} X.-H. Chen, Y.-H. Zhai, D. Zhang, and L.-A. Wu,``Absolute self-calibration of the quantum efficiency of single-photon detectors,''
\href{https://doi.org/10.1364/OL.31.002441}{\textit{Opt. Lett.} \textbf{31} 2441 (2006).} 

\bibitem{Lopez} M. L\'{o}pez, H. Hofer, and S. K\"{u}ck,
 ``Detection efficiency calibration of single-photon silicon avalanche photodiodes traceable using double attenuator technique,''
\href{http://dx.doi.org/10.1080/09500340.2015.1021724}{\textit{J. Mod. Opt.} \textbf{62} (2015).} 

\bibitem{Dovrat} L. Dovrat, M. Bakstein, D. Istrati, A. Shaham, and H. S. Eisenberg, ``Measurements of the dependence of the photon-number distribution on the number of modes in parametric down-conversion,'' \href{https://doi.org/10.1364/OE.20.002266}{\textit{Opt. Exp.} \textbf{20}, 2266 (2010).}

\bibitem{DovratSim} L. Dovrat, M. Bakstein, D. Istrati, and H. S. Eisenberg, ``Simulations of photon detection in silicon photomultiplier number-resolving detectors,'' \href{http://dx.doi.org/10.1088/0031-8949/2012/T147/014010}{\textit{Phys. Scr.} \textbf{T147}, 014010 (2012).}

\bibitem{saleh} B.E.A Saleh and M. C. Teich, \textit{Fundamentals of photonics}, New York: Wiley  (1991) 

\bibitem{Paul96} H. Paul, P. T\"{o}rm\"{a}, T. Kiss, and I. Jex, ``Photon Chopping: New Way to Measure the Quantum State of Light,'' \href{https://doi.org/10.1103/PhysRevLett.76.2464}{\textit{Phys. Rev. Lett.} \textbf{76}, 2464 (1996).}

\bibitem{wolfram}  Eric W. Weisstein, ``Stirling Number of the Second Kind.'' \href{http://mathworld.wolfram.com/StirlingNumberoftheSecondKind.html}{\textit{Wolfram MathWorld}}

\bibitem{Buzhan} P. Buzhan, B. Dolgoshein, L. Filatov, A. Ilyin, V. Kaplin, A. Karakash, S. Klemin, R. Mirzoyan, A. Otte,
E. Popova, V. Sosnovtsev, and M. Teshima, ``Large area silicon photonmultipliers: Performance and applications,''
\href{http://dx.doi.org/10.1016/j.nima.2006.05.072}{\textit{Nucl. Instrum. Methods} \textbf{567}, 78 (2006).}

\bibitem{Afek} I. Afek, A. Natan, O. Ambar, and Y. Silberberg, ``Quantum state measurements using multipixel photon detectors,'' \href{http://dx.doi.org/10.1103/PhysRevA.79.043830}{\textit{Phys. Rev. A} \textbf{79}, 043830 (2009).}

\bibitem{Akiba} M. Akiba, K. Tsujino, K. Sato, and M. Sasaki, ``Multipixel silicon avalanche photodiode with ultralow dark count rate at liquid nitrogen temperature,''  \href{https://doi.org/10.1364/OE.17.016885}{\textit{Opt. Exp.} \textbf{17}, 16885 (2009).}

\bibitem{Eraerds} P. Eraerds, M. Legr\'{e}, A. Rochas, H. Zbinden, and N. Gisin, ``SiPM for fast Photon-Counting and Multiphoton Detection,'' 
\href{https://doi.org/10.1364/OE.15.014539}{\textit{Opt. Exp.} \textbf{15}, 14539 (2007).}


\end{thebibliography}
\end{document}